\documentclass[11pt, a4paper]{article}
\pdfoutput=1

\usepackage{anysize}
\marginsize{3cm}{3cm}{1.5cm}{1.5cm}
\usepackage[T1]{fontenc}
\usepackage{inputenc}

\usepackage{amsfonts}
\usepackage{amssymb}
\usepackage{graphics}
\usepackage{epsfig}
\usepackage{graphicx}
\usepackage{epsfig}
\usepackage{amssymb}
\usepackage{amsfonts}

\usepackage{amsmath}
\usepackage{xcolor}
\usepackage{float}
\usepackage{enumerate}
\usepackage{slashed}
\usepackage{hyperref}
\hypersetup{colorlinks=true,linkcolor=blue,citecolor=blue} 

\usepackage{caption}
\usepackage{subcaption}

\numberwithin{equation}{section}

\newcommand {\beq} {\begin{equation}}
\newcommand {\eeq} {\end{equation}}
\newcommand{\bea}{\begin{eqnarray}}
\newcommand{\eea}{\end{eqnarray}}
\newcommand{\bit}{\begin{itemize}}
\newcommand{\eit}{\end{itemize}}
\def\nl{\nonumber \\}

\def\sgn{{\rm  sgn}}

\def\a{\alpha}

\def\b{\beta}

\def\p{\partial}

\def\le{\left(}
\def\ri{\right)}

\def\beq{\begin{equation}}
\def\eeq{\end{equation}}

\begin{document}

\begin{titlepage}

\begin{flushright}

\end{flushright}
\bigskip
\begin{center}
{\LARGE  {\bf
Subsystem complexity in warped AdS
  \\[2mm] } }
\end{center}
\bigskip
\begin{center}
{\large \bf  Roberto  Auzzi$^{a,b}$},
 {\large \bf Stefano Baiguera$^{c}$},
  {\large \bf Arpita Mitra$^{e}$}, \\
   {\large \bf Giuseppe Nardelli$^{a,d}$} 
 {\large \bf and }     {\large \bf Nicol\`o Zenoni$^{a,b}$}
\vskip 0.20cm
\end{center}
\vskip 0.20cm 
\begin{center}
$^a${ \it \small  Dipartimento di Matematica e Fisica,  Universit\`a Cattolica
del Sacro Cuore, \\
Via Musei 41, 25121 Brescia, Italy}
\\ \vskip 0.20cm 
$^b${ \it \small{INFN Sezione di Perugia,  Via A. Pascoli, 06123 Perugia, Italy}}
\\ \vskip 0.20cm 
$^c${ \it \small{Universit\`a degli studi di Milano Bicocca and INFN, 
Sezione di Milano - Bicocca, \\ Piazza
della Scienza 3, 20161, Milano, Italy}}
\\ \vskip 0.20cm 
$^d${ \it \small{TIFPA - INFN, c/o Dipartimento di Fisica, Universit\`a di Trento, \\ 38123 Povo (TN), Italy} }
\\ \vskip 0.20cm 
$^e${ \it \small{ Indian Institute of Science Education and Research Bhopal, \\ Bhopal Bypass, Bhauri, Bhopal 462006, India} }
\\ \vskip 0.20cm 
E-mails: roberto.auzzi@unicatt.it, s.baiguera@campus.unimib.it,  \\
 arpitam@iiserb.ac.in, giuseppe.nardelli@unicatt.it, nicolo.zenoni@unicatt.it  \end{center}
\vspace{3mm}

\begin{abstract}

We compute the ultraviolet divergences 
of holographic subregion complexity 
for the left and right factors of the thermofield double state in
warped AdS$_3$ black holes, both for the action and the volume conjectures.
Besides the linear divergences, which are also present in the BTZ black hole, 
additional logarithmic divergences appear.
For the action conjecture,  these log divergences 
are not affected by the arbitrarity  in the length scale
associated with the counterterm needed  to ensure reparameterization invariance.
We find that the subregion action complexity
obeys the superadditivity property for the thermofield double
in warped AdS$_3$, independently from the action counterterm coefficient.
 We study the temperature dependence of subregion complexity  at constant angular momentum
 and we find that it is correlated with the sign of the specific heat.

\end{abstract}

\end{titlepage}

\section{Introduction}

According to the AdS/CFT correspondence, 
eternal black holes in AdS are dual
to the thermofield double state \cite{Maldacena:2001kr,Hartman:2013qma},  
which corresponds to two copies of entangled conformal field theories
 living on the left and right boundaries of the Penrose diagram. 
Quantum information concepts such as entanglement entropy have the potential
to give us essential information on how spacetime can emerge from the boundary field theory
in holographic dualities.  

Recently,  the new concept of computational complexity
has been introduced in order to provide a field theory dual to the linear growth
of the Einstein-Rosen Bridge inside the horizon of a Black Hole (BH)
\cite{Susskind:2014rva,Stanford:2014jda,Susskind:2014moa}.
The concept of complexity originates in theoretical computer science.
In quantum computing, it is defined as the minimal number of elementary
unitary operations which are needed in order to prepare a given state
from a reference state.  In quantum mechanics, this is a function
of the chosen reference state, of the set of of quantum gates
and of the allowed tolerance in the accuracy with which the final state is prepared.
An elegant geometric approach to complexity 
was developed by Nielsen and collaborators \cite{Nielsen1,Nielsen2}.
In conformal field theories (CFTs), a precise definition of complexity
is still lacking.  Some possible definitions for free field theories
have been studied by several authors, e.g.
  \cite{Jefferson:2017sdb,Chapman:2017rqy,Hashimoto:2017fga,Kim:2017qrq,Khan:2018rzm,Hackl:2018ptj,Chapman:2018hou}.
Another approach for 2-dimensional field theories uses the Liouville action
\cite{Caputa:2017yrh,Bhattacharyya:2018wym}.
This research field is still in its nascent stages.

There are a few proposals for the gravitational dual of complexity in the AdS/CFT correspondence:
\begin{itemize}
\item
 the Complexity=Volume (CV) \cite{Susskind:2014rva,Stanford:2014jda,Susskind:2014moa}
  refers to the spacetime volume $V$ of an extremal spacelike codimension-$1$ slice anchored at the boundary.
  We will denote the corresponding complexity as $\mathcal{C}_V$:
  \beq
\mathcal{C}_V = \, {\rm Max}  \le \frac{V}{G l} \ri \, ,
\eeq
where $G$ is the gravitational constant and $l$ the AdS radius.
 \item
 the Complexity=Action (CA) \cite{Brown:2015bva,Brown:2015lvg}
  refers to a gravitational action $I_{WDW}$ computed in the Wheeler-De Witt (WDW) patch, 
  which is the union of all the spacelike slices that can be attached to the boundary
  at a given time.
  The gravitational action must include also a set of boundary terms needed to consistently
 define an additive and a reparameterization-invariant bulk action.
 This includes the Gibbons-Hawking-York term, the terms due to the presence of null boundaries 
 and their intersections  \cite{Hayward:1993my,Lehner:2016vdi}.
 We denote the action complexity as $\mathcal{C}_A$:
 \beq
\mathcal{C}_A=  \frac{I_{WDW}}{\pi } \, .
\eeq
\item
  the Complexity= Spacetime Volume (CV 2.0)  \cite{Couch:2016exn}
  refers to the spacetime volume of the WDW patch:
   \beq
\mathcal{C}_V^{\rm 2.0} \propto  V_{WDW} \, .
\eeq
  \end{itemize}
These proposals share several common qualitative behaviours.
Our understanding of complexity  in field theory is not good enough  at the moment to precisely
discriminate among them. CA appears as more universal, because
 the CV proposal requires the introduction of an \emph{ad hoc} length scale to relate complexity, 
 which is dimensionless, to a spacetime volume, which is a dimensional quantity. 
On the other hand,  CV shows a more regular monotonically increasing growth rate for intermediate times
\cite{Carmi:2017jqz}, which matches expectations from quantum circuits.
See e.g. \cite{Yang:2018nda,Hashimoto:2018bmb,Chapman:2016hwi,Cai:2016xho,Reynolds:2016rvl,
Moosa:2017yvt,Moosa:2017yiz,Chapman:2018dem,Chapman:2018lsv,Barbon:2015ria,Bolognesi:2018ion,Swingle:2017zcd,An:2018xhv}
for several recent investigations on the topic of holographic complexity.

By analogy with entanglement entropy,
an interesting further extension of the CV and CA conjectures 
is to restrict to subregions of the full boundary theory.  
This should  be dual to some notion of
 subsystem complexity of a mixed state on the boundary.
The most promising bulk region which should correspond to 
a boundary density matrix localised in a subregion
is the entanglement wedge \cite{Headrick:2014cta}, and so it is natural
to propose that subregion complexity is dual to quantities which have support
in the entanglement wedge. For CV, it was conjectured \cite{Alishahiha:2015rta}
that mixed state complexity is dual to the volume of the codimension-$1$ extremal slice in the bulk 
attached to the boundary subregion and its Ryu-Takayanagi (RT) \cite{Ryu:2006bv} surface.
In the CA framework, it was proposed in \cite{Carmi:2016wjl}
that the mixed state complexity is dual to the action of the intersection of
the WDW patch and the entanglement wedge  associated 
to the  given spatial subregion. 
Other works on holographic subregion complexity include 
\cite{Ben-Ami:2016qex,Abt:2017pmf,Abt:2018ywl,Roy:2017kha,Roy:2017uar,Bhattacharya:2019zkb}.

There are  a few different possible 
 definitions of complexity of a  mixed state $\rho$ localised in a subregion 
  of the Hilbert space of a quantum field theory \cite{Agon:2018zso}:
 \begin{itemize}
 \item  purification complexity $\mathcal{C}_P$, which 
 can be defined as the minimal number of gates needed to transform the initial
 pure state (plus some ancillary external qubits) into a purification of the  
 mixed state $\rho$;
\item spectrum complexity $\mathcal{C}_S$, which can be defined as the minimal number
of operations needed to prepare a mixed state $\rho_{spec}$ with the same spectrum as $\rho$;
\item basis complexity $\mathcal{C}_B$, which can be defined as the minimum number of gates
needed to prepare   $\rho$ from $\rho_{spec}$.
 \end{itemize}
The spectrum complexity does not reduce to 
complexity when computed on pure states, and so it is not a good candidate
 as a field theory dual of CV or CA. Instead both $\mathcal{C}_P$ and $\mathcal{C}_B$ might be
 in principle  reasonable candidates as duals of holographic complexities.
These issues were recently investigated by
 \cite{Agon:2018zso,Alishahiha:2018lfv,Caceres:2018blh,Ghodrati:2019hnn}.

In particular, it was conjectured  \cite{Agon:2018zso} that $\mathcal{C}_P$  should be subadditive for the left $L$
and right $R$ factors of the thermofield double state $TD$.
 An analog guess was made about superadditivity of $\mathcal{C}_B$.
If these conjectures were true, they would be useful to discriminate 
which notion of subregion complexity is dual to a given holographic 
realisation. 

The volume complexity $\mathcal{C}_V$ is in general superadditive
because the volume is always a positive-definite quantity:
\beq
 \mathcal{C}_V( AB)  \geq  \mathcal{C}_V( A) +  \mathcal{C}_V( B)  \, .
\eeq
Moreover, for the thermofield double at time zero, this inequality saturates:
\beq
 \mathcal{C}_V( TD, t=0) =  \mathcal{C}_V( L) +  \mathcal{C}_V( R) \, .
 \label{saturation}
\eeq
For $\mathcal{C}_A$ the situation is murky, because action is not positive-definite.
 An interesting technical point which arises  in the CA conjecture
 is due to an arbitrary length  scale $\tilde{L}$ which appears in
a counterterm which is needed in order 
to make the action reparameterization invariant \cite{Lehner:2016vdi}.
Depending on the choice of $\tilde{L}$, for the AdS neutral black hole,
one can get either \cite{Alishahiha:2018lfv}
that CA is superadditive or subadditive for the $L,R$ sides of the thermofield  double. 

Another interesting property is the temperature behaviour
of mixed state complexity. In \cite{Agon:2018zso} 
it was argued from tensor network arguments
that $\mathcal{C}_B$ decreases with temperature $T$ and approaches zero
for $T \rightarrow \infty$, while $\mathcal{C}_P$ should not have strong dependence on $T$.
As studied in \cite{Agon:2018zso,Alishahiha:2018lfv},  for the AdS neutral black hole
the behaviour of subsystem action complexity as a function of temperature 
also depends on $\tilde{L}$.

An interesting ultraviolet modification of the AdS/CFT correspondence
in which many results of holography can be generalised  and extended
is  the Warped AdS$_3$/WCFT$_2$ correspondence 
\cite{Anninos:2008fx,Anninos:2008qb,Detournay:2012pc,Hofman:2014loa,Jensen:2017tnb}.
This is a correspondence between  gravitational bulk theories in $2+1$ dimensions
in a spacetime with Warped AdS$_3$ (WAdS$_3$) asymptotic metric and a class of non-relativistic
theories in $1+1$ dimensions on the boundary. These are called Warped Conformal Field Theories 
(WCFTs), and are invariant under the Virasoro and the $U(1)$ Kac-Moody current algebras.
They provide a natural direction to extend  holography in a non-relativistic direction.
For example, a Cardy formula reproducing the black hole entropy
 was derived in \cite{Detournay:2012pc}. Entanglement entropy  
has been studied by several authors \cite{Anninos:2013nja, Castro:2015csg, Azeyanagi:2018har,Song:2016pwx,Song:2016gtd}. 
Complexity was investigated in \cite{Ghodrati:2017roz,Auzzi:2018zdu,Auzzi:2018pbc,Dimov:2019fxp}.

In this work we compute the divergences of subregion 
complexity for the left and right factors of the thermofield double state,
 in the case of black holes  in asymptotically WAdS$_3$ spacetimes.
We  investigate the temperature dependence of subregion complexity in each of the conjectures
and the sub/superadditivity properties of the CA conjecture.
These properties may help to discriminate which notion of subregion
complexity (for example $\mathcal{C}_P$ or $\mathcal{C}_B$) is  dual to each 
of the holographic complexity conjectures. 

We find several features that differ from the AdS case:
\begin{itemize}
\item the structure of divergences of subregion complexity  is different from the AdS case:
besides the linear term in the cutoff $\Lambda$,
an additional $\log \Lambda$ divergence arises.
\item subregion $\mathcal{C}_A$ is always superadditive for the $L$, $R$ factors of TD.
Instead in AdS the  sub/superadditivity property is dependent on 
the arbitrary parameter $\tilde{L}$.
\item subregion complexities have a temperature dependence
which is correlated with specific heat.
\end{itemize}

The paper is organized as follows:
in section \ref{warped-bh} we review some basic properties of
warped black holes realised in Einstein gravity. In section \ref{azione}
we compute the divergences of total and subregion action for
rotating black holes. In section \ref{volume} we compute these divergences 
in CV and CV 2.0. We conclude and discuss our results in section
\ref{conclu}. The details of the calculation for the non-rotating case and some
other technical details are deferred to the appendices.


\section{Black holes in warped $ \mathrm{AdS}_3 $ spacetime}
\label{warped-bh}

Black holes in asymptotically warped $ \mathrm{AdS}_3 $ spacetime
 \cite{Anninos:2008fx,Moussa:2003fc,Bouchareb:2007yx} are described by the following metric
\beq
\frac{ds^2}{l^2} = dt^2 + \frac{dr^2}{(\nu^2 +3)(r-r_+)(r-r_-)} + \le 2 \nu r - \sqrt{r_+ r_- (\nu^2 +3)} \ri dt d \theta + \frac{r}{4} \Psi (r) d \theta^2 \, ,
\label{metric warped BTZ general}
\eeq
where 
\beq
\Psi (r)= 3 (\nu^2 -1) r + (\nu^2 +3) (r_+ + r_-) - 4 \nu \sqrt{r_+ r_- (\nu^2 +3)} \, ,
\eeq
the inner and outer horizons are placed in $ r_-, r_+ $ and $ \nu $
 is a warping parameter such that for $ \nu=1 $ the metric gives the BTZ black hole \cite{Banados:1992wn,Banados:1992gq}.
We define $ \rho_0 $ as the zero of the function $ \Psi (r) , $ \emph{i.e.}:
\beq
\rho_0 =  \frac{4 \nu \sqrt{r_+ r_- (\nu^2 +3)} - (\nu^2 +3)(r_+ + r_-)}{3(\nu^2 -1)} \, .
\eeq
We introduce $\tilde{r}_0$, defined by
\beq
\tilde{r}_0 = \max \le 0, \rho_0 \ri \, ,
\eeq
so that the range of variables is: $\tilde{r}_0 \leq r < \infty$, $- \infty < t < \infty$ and $\theta \sim \theta + 2 \pi$.

This metric is pathologic when $ \nu^2 < 1, $ because admits closed timelike curves.
Temperature and angular velocity of the outer horizon are \cite{Anninos:2008fx}:
\beq
T= \frac{\nu^2+3}{4 \pi l } \,  \frac{r_+ -r_-}{2 \nu r_+ -\sqrt{(\nu^2+3) r_+ r_-} } \, ,
\qquad
\Omega=\frac{2}{(2 \nu r_+ -\sqrt{(\nu^2+3) r_+ r_-}) l } \, .
\label{TT}
\eeq

Entropy, mass and angular momentum depend on the gravitational action we choose, 
and for our computations we will consider warped BHs arising 
as a solution of Einstein gravity plus Maxwell and Chern-Simons terms  \cite{Banados:2005da,Barnich:2005kq}.
The entropy is then proportional to the area of the event horizon
\beq
S=\frac{l \pi}{4 G} (2 \nu r_+ - \sqrt{r_+ r_- (\nu^2+3)}) \, ,
\eeq
and the conserved charges (mass and angular momentum) are \cite{Banados:2005da,Barnich:2005kq}:
\beq
M=\frac{1}{16 G} (\nu^2+3)
 \le  \le r_{-} + r_{+} \ri - \frac{\sqrt{r_{+}r_{-} (\nu^2 +3)}}{\nu} \ri \, ,
 \label{MM}
\eeq
\beq
J=\frac{l}{32 G} (\nu^2+3) 
\le \frac{r_- r_+ (3+5 \nu^2)}{2 \nu}
 -(r_+ + r_-) \sqrt{(3+\nu^2) r_+ r_-} 
 \ri \, .
\label{JJ}
\eeq

\subsection{An explicit realization in Einstein gravity}

The solution  in eq. (\ref{metric warped BTZ general})
can be obtained as a vacuum solution of Topological Massive Gravity
 \cite{Anninos:2008fx,Moussa:2003fc,Bouchareb:2007yx}
 or  New Massive Gravity \cite{Clement:2009gq}.
 We will instead focus on realizations of  the metric (\ref{metric warped BTZ general})
 in Einstein gravity. Unfortunately, all the known explicit constructions
 of WAdS$_3$ black holes in Einstein gravity have some pathology in the matter content.
 For concreteness we will use a model introduced in \cite{Banados:2005da},
 where the matter content is a Chern-Simons $U(1)$ gauge field.
 In order to find absence of closed timelike curves ($\nu^2 \geq 1$),
 a ghost-like kinetic Maxwell term is needed. This is the same theoretical
 setting that was studied in \cite{Auzzi:2018pbc}, where we found that 
 the asymptotic growth of $\mathcal{C}_A$ was, as expected, proportional to $T S$.
 This is also consistent with the CV conjecture \cite{Auzzi:2018zdu}.
 The CA conjecture seems solid enough to survive to unphysical
 matter contents which include ghosts.

We consider  Einstein gravity in $3$ dimensions with a negative cosmological constant, coupled
to a $U(1)$ gauge field with both Maxwell and Chern-Simons terms 
\cite{Banados:2005da}:
\beq
I_{\mathcal{V}} = \frac{1}{16 \pi G} \int_{\mathcal{V}} d^3 x \, \left\lbrace \sqrt{-g} \left[ \le R + \frac{2}{L^2} \ri - \frac{\kappa}{4} F_{\mu \nu} F^{\mu\nu} \right] - \frac{\alpha}{2} \epsilon^{\mu\nu\rho} A_{\mu} F_{\nu \rho} \right\rbrace = \int_{\mathcal{V}} d^3 x \, \sqrt{-g} \, \mathcal{L} \, ,
\label{bulk action}
\eeq
We use the same notation as in \cite{Auzzi:2018pbc}.
The following solution for the gauge field is considered:
\beq
A= a \, dt +  c r \,  d \theta \, , \qquad
F = c \, dr \wedge d \theta \, ,
\eeq
The Maxwell and Einstein equations  give:
\beq
\a = \kappa \frac{\nu}{l} \, ,
\qquad
L = l \sqrt{\frac{2}{3-\nu^2} } \, , \qquad
c =\pm  l \sqrt{\frac{3}{2 } \frac{1-\nu^2}{\kappa}}
\, . 
\label{gaugau}
\eeq
In order to avoid  closed timelike curves, we have to impose
$\nu \geq 1$, which implies the ghost condition $\kappa<0$.

The gauge parameter $a$ is not constrained by the equations of motion,
but the action depends explicitly on $a$ through the Chern-Simons term.
The value of $a$ is important to properly define 
the mass $M$ as a conserved charge  \cite{Barnich:2005kq}.
Formally, only for the value
\beq
a=
\frac{l}{\nu} \sqrt{\frac32} \sqrt{\nu^2-1}  \, .
\label{a-action}
\eeq
the mass is associated to the
Killing vector $\p/\p t$ and it does not depend
 on the $U(1)$ gauge transformations. 
For this value, the action density reads:
\beq
16 \pi G \, \sqrt{-g} \, \mathcal{L} = - \frac{l}{2} (\nu^2 +3) = \mathcal{I} \, .
\label{densita}
\eeq

\subsection{Eddington-Finkelstein coordinates}

It is convenient to introduce null coordinates using the ADM decomposition of the metric 
(\ref{metric warped BTZ general}):
 \beq
 \label{BHADM}
 ds^2=-N^2 dt^2  +\frac{l^4 dr^2}{4 R^2 N^2}
 +l^2 R^2 (d \theta + N^\theta dt)^2
 \, , 
 \eeq
 where
 \beq
 R^2=\frac{r}{4} \Psi \, , \qquad
 N^2=\frac{l^2 (\nu^2+3) (r-r_+)(r-r_-)}{4 R^2} \, ,
 \qquad
 N^\theta= \frac{2 \nu r -\sqrt{r_+ r_- (\nu^2+3)} }{2 R^2} \, .
  \label{ADM quantities}
 \eeq
In order to delimit the WDW patch, we use the set of null coordinates
 introduced in  \cite{Jugeau:2010nq}, obtained by considering 
 a set of null geodesics  which satisfies $(d \theta + N^\theta dt)=0$.
 Then the last term in the metric (\ref{BHADM})
 saturates to zero, and the null geodesics are parameterized by the constant
  $u$ and $v$ trajectories:
 \beq
du = u_{\alpha} dx^{\alpha} = dt - \frac{l^2}{2 R N^2} dr \, , \qquad
dv = v_{\alpha} dx^{\alpha} =dt + \frac{l^2}{2 R N^2} dr  \,  .
\label{one forms null directions}
 \eeq
These one-forms are both normal and tangent to the null boundaries of the WDW patch.
Moreover, the integral curves of $u^\a$ and $v^\a$ are 
null geodesics in the affine parameterization, i.e.
\beq
u^\a D_\a u^\b=0 \, , \qquad v^\a D_\a v^\b=0 \, ,
\label{affine}
\eeq
where $D_\a$ is the covariant derivative.

From eq. (\ref{one forms null directions}), we find the null coordinates
\beq
u= t - r^* (r) \, , \qquad
v= t + r^* (r) \, ,  
\label{Kruskalcoordinates}
\eeq
where the tortoise coordinate $r^*$ is given by
\beq
r^*(r)= \int^r \frac{dr'}{f(r')} \, , \qquad    f(r)=\frac{2 R N^2}{l^2} 
=\frac{(\nu^2 +3)(r-r_-)(r-r_+)}{\sqrt{r \Psi(r)}} \, .
\label{derirstar}
\eeq 
Integrating eq. (\ref{derirstar}),  $r^*$ can be explicitly found  \cite{Jugeau:2010nq};
for $r_+ \neq r_-$ the explicit expression is
\beq
\begin{aligned}
r^{*}(r)= & \frac{\sqrt{3 \left(\nu^2-1\right)}}{\left(\nu^2+3\right)} 
\left\lbrace \frac{\sqrt{r_{+}(r_{+}-\rho_0)}}{r_+-r_-} \log \left(\frac{\left|r-r_{+}\right|}{\left(\sqrt{r} \sqrt{r_{+}-\rho_0}+\sqrt{r-\rho_0}\sqrt{r_+}\right)^2}\right) \right.\\
& \left. - \frac{\sqrt{r_{-}(r_{-}-\rho_0)}}{r_+-r_-} \log \left(\frac{\left|r-r_{-}\right|}{\left(\sqrt{r} \sqrt{r_{-}-\rho_0}+\sqrt{r-\rho_0}\sqrt{r_-}\right)^2}\right)  +2\log(\sqrt{r} +\sqrt{r-\rho_{0}}) \right\rbrace \, .
\label{rstar}
\end{aligned}
\eeq
The tortoise coordinate $r^*$ is divergent at $ r \rightarrow \infty$, with leading behaviour
\beq
\lim_{r \rightarrow \infty} r^* (r) \approx \frac{\sqrt{3(\nu^2 -1)}}{\nu^2 +3} \log r \, .
\eeq 
The non-rotating case is defined by $J=0$, 
and  is realised by one of the following conditions:
\beq
r_-=0 \, , \qquad  \frac{r_+}{r_-}=\frac{4 \nu^2}{\nu^2+3} \, .
\label{non-rota}
\eeq
The corresponding Penrose diagram is the same as the one for Schwarzschild BH
in four dimensions  \cite{Jugeau:2010nq}. 
 The two values in eq (\ref{non-rota}) can be mapped one into the other by an isometry \cite{Jugeau:2010nq},
 and so for  simplicity we will always consider the case $ r_-=0, r_+ = r_h$ 
 when we will refer to the non-rotating case.

For generic values $(r_+,r_-)$ we get a rotating BH, 
and the Penrose diagram is  the same as the one of the
Reissner-Nordstr\"om BH.
The extremal limit corresponds to 
$r_+ = r_- $; in this case the temperature is zero
and there is no thermofield double: the Penrose diagram 
has just one boundary. We are not interested in this case
in the present paper.

\section{Action}
\label{azione}

\subsection{Contributions to the action}

The action of the WDW patch has several contributions, which can be
evaluated using the results of \cite{Lehner:2016vdi}:
\beq
I = I_{\mathcal{V}} + I_{\mathcal{B}} + I_{\mathcal{J}} + I_{\rm ct} \, .
\eeq
Here $ I_{\mathcal{V}} $ refers to the bulk, $ I_{\mathcal{B}} $ to the codimension-1 boundaries 
and $ I_{\mathcal{J}} $ to the codimension-2 joints coming from intersections of other boundaries.
The contribution $ I_{\rm ct} $ is a counterterm to be added  in order to ensure
 reparameterization invariance  of the action.

We are interested in the divergent parts of complexity,
so we introduce an UV cutoff at $ r= \Lambda$.
 We will focus on the thermofield double state at time zero, i.e.
\beq
t_L = t_R = t_b= 0 \, ,
\eeq
which by symmetry  corresponds to the minimum of the action.
The time dependence of the finite part of complexity was previously studied in \cite{Auzzi:2018pbc}.

\begin{figure}[h]
\centering
\begin{subfigure}[c]{0.4\linewidth}
\includegraphics[scale=0.53]{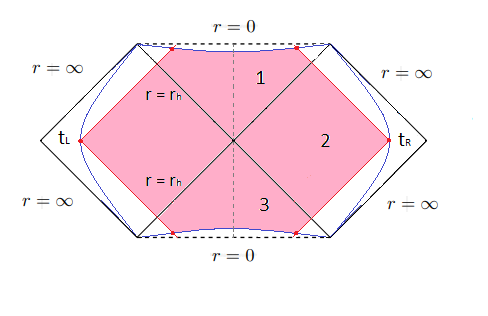}
\end{subfigure} \hspace{15mm}
\begin{subfigure}[c]{0.4\linewidth}
\includegraphics[scale=0.53]{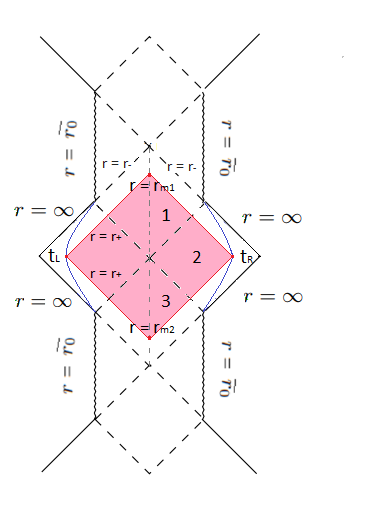}
\end{subfigure}
\caption{Penrose diagram and WDW patch at $t_b=0$ for the non-rotating (left)
and rotating (right) black holes.}
\label{WDW}
\end{figure}

The Penrose diagram and the corresponding WDW patch in the non-rotating 
case are depicted on the left side of Fig. \ref{WDW}.
The configuration of the WDW patch in the rotating case is depicted 
on the right side of Fig. \ref{WDW}.
We call $  r_{m1}, r_{m2} $ the null joints referring respectively to the top and bottom vertices of the spacetime region of interest.
The definition of the null joints in terms of the tortoise coordinates are
\beq
\frac{t_b}{2} + r^*_{\Lambda} - r^* (r_{m1} ) = 0 \, , \qquad
\frac{t_b}{2} - r^*_{\Lambda} + r^* (r_{m2} ) = 0 \, ,
\label{definizione dei joint}
\eeq
where $r^*_{\Lambda}=r^*(\Lambda)$. 

{\bf Bulk contributions:} We follow the calculation in \cite{Auzzi:2018pbc}.
The integrand of the bulk action is constant, 
and so this contribution is proportional to the spacetime volume enclosed in the WDW patch.
It is convenient to separate this bulk region into three parts, as indicated 
in Fig. \ref{WDW}  for the two cases:
\beq
I^{1}_{\mathcal{V}}  = \frac{\mathcal{I}}{8 G} \int_{\bar{r}_1}^{r_+} dr \le \frac{t_b}{2} + r^{*}_{\Lambda} - r^* (r) \ri \, ,
\qquad \, 
I^{2}_{\mathcal{V}}  = \frac{\mathcal{I}}{4 G} \int_{r_+}^{\Lambda} dr \le  r^{*}_{\Lambda} - r^* (r) \ri \, ,
\label{bulk12}
\eeq
\beq
I^{3}_{\mathcal{V}}  = \frac{\mathcal{I}}{8 G} \int_{\bar{r}_2}^{r_+} dr \le - \frac{t_b}{2} + r^{*}_{\Lambda} - r^* (r) \ri \, ,
\label{bulk3}
\eeq
where $\mathcal{I}$ is defined in eq. (\ref{densita})
and the angular part  gives a factor of $ 2 \pi$.
The integration range in the non-rotating case is done up to a cutoff $\epsilon_0 \approx 0$:
\beq
\bar{r}_1 =\bar{r}_2= \varepsilon_0 \, , 
\eeq 
 and in the rotating case is:
\beq
\bar{r}_1 = r_{m1} \, , \qquad
\bar{r}_2 = r_{m2}  \, .
\eeq 

{ \bf Boundary terms:}
In the non-rotating case, we have to include the Gibbons-Hawking-York (GHY) term
for the two spacelike boundaries:
\beq
I_{\mathcal{B}} =  - \frac{ 1 }{8 \pi G}  \int_{\mathcal{B}} d^2 x \, \sqrt{|h|} \, K \, ,
\label{gibb-hawk}
\eeq
where $ \mathcal{B} $ is the boundary, \emph{h} the induced metric determinant, \emph{K} the trace of the extrinsic curvature.
Both in the rotating and non-rotating cases, we have in principle contributions
from null boundaries that we set to zero by using an affine parameterization (see eq. (\ref{affine}))
for the boundary geodesics \cite{Lehner:2016vdi}.

{\bf Joint terms:}
Given a joint $ \Sigma $ where various boundaries meet, the contribution to the action is given by
\beq
I_{\mathcal{J}} = \frac{1}{8 \pi G} \int_{\Sigma} d \theta \, \sqrt{\sigma} \mathfrak{a} \, ,
\label{jjj1}
\eeq
where $ \sigma $ is the determinant of the induced metric on the joint (in this case, it is 1-dimensional):
\beq
\sqrt{\sigma} = l \, R (r) = l \sqrt{\frac{r}{4} \Psi (r)} \, .
\label{misu}
\eeq
The integrand $ \mathfrak{a} $ depends on the dot product of normal one-forms $ \mathbf{k}_L , \mathbf{k}_R $
 defined on the boundaries meeting at the joint.

All the joints appearing in Fig. \ref{WDW} arise from intersections of null lines apart from the ones located at the past and future singularities.
However, the latter vanish when we send the IR cutoff to 0. 
We will thus focus only on joints where two null lines meet, in which case the integrand is given by
\beq
\mathfrak{a} = \eta \log \left| \frac12 \mathbf{k}_L \cdot \mathbf{k}_R \right| \, ,
\eeq
where $ \eta= \pm 1 $ depending on the position of the joint with respect to the future direction of time and the location with respect to the interior of the WDW patch \cite{Lehner:2016vdi}.
In our case,  the joints in the interior of the black and white holes
in figure \ref{WDW} have $\eta=1$, while the ones nearby the UV cutoff
have $\eta=-1$. The null directions are parameterized 
according to eq. (\ref{one forms null directions}), which gives
\beq
\mathfrak{a}= \eta \log \left| \frac12 u^{\alpha} v_{\alpha} \right| = 
\eta \log \left| \frac{A^2}{l^2} \frac{2 R(r)}{f(r)} \right|
 \, ,
\eeq
where $A$ is an overall arbitrary constant 
which parameterizes the ambiguity in defining the null normals on the boundaries.
From eq.  (\ref{jjj1}), we find a general expression for all the null joints in the WDW patch
\beq
I_{\mathcal{J}} = - \sum_k \eta_k \frac{l}{4G} \sqrt{\frac{r_k}{4} \Psi (r_k)} \log \left| \frac{l^2}{A^2} \frac{f(r_k)}{2 R(r_k)} \right| \, ,
\label{general formula joints}
\eeq
where the sum is  over the $k$ joints, whose radial coordinate is $ r_k $.

{\bf Counterterm for the null boundaries:}
The following counterterm  \cite{Lehner:2016vdi} must be added to the 
boundary term of null boundaries, in order to make the action 
invariant under reparameterization: 
\beq
I_{\rm ct}=\frac{1}{8\pi G}\int d\theta \, d\lambda \, \sqrt{\sigma} \, \Theta\log{|\tilde{L}\Theta|} \, ,
\label{counterterm}
\eeq
where  $\lambda$ is the affine parameter of the null geodesics which delimit the boundary, and
\beq
\Theta=\frac{1}{\sqrt{\sigma}}\frac{\partial\sqrt{\sigma}}{\partial \lambda} \, ,
\eeq
is the expansion of the congruence of null geodesics on the hypersurface.
The parameter $\tilde{L}$ appearing in eq. (\ref{counterterm}) is an arbitrary length scale
which is needed for dimensional reasons, whose physical meaning is so far obscure.
  
We will need to evaluate the counterterm along the null lines described by the null coordinates $(u,v) .$
It is convenient to re-express the integral (\ref{counterterm})
in terms of  $r$, using
\beq
\frac{\p r}{\p \lambda} = A  v^r 
= \frac{2A}{l^2}  R (r) \, .
\eeq
Thus, eq. (\ref{counterterm}) becomes
\beq
I_{\rm ct}
 = \frac{1}{4G} \int_{r_{\rm inf}}^{r_{\rm sup}} dr \, \frac{\p \sqrt{\sigma}}{\p r} 
 \log \left| \frac{\tilde{L}}{\sqrt{\sigma}} \frac{\p \sqrt{\sigma}}{\p r} \frac{2A}{l^2}  R (r) \right| \, ,
 \label{controtermineA}
\eeq
where we integrate between the endpoints along the radial direction of the null line.

\subsection{Total action}

In this section we compute the divergences of the total action
 of the WDW patch at $t_b=0$ in the rotating case.
The calculation for the non-rotating case involves slightly different details
which are sketched in Appendix \ref{appe-total};
as expected, the  result reproduces the $r_- \rightarrow 0$ limit of the rotating case.
The Penrose diagram in the rotating case is depicted in Fig. \ref{WDW}, 
where the top and bottom joint vertices are described by the expressions (\ref{definizione dei joint}).
At  $ t_b =0$,  we get
\beq
r^*_{\Lambda} = r^* (r_{m1} ) = r^* (r_{m2} ) \equiv r^* (r_m) \, ,
\label{top and bottom joints at t=0}
\eeq
and the configuration is symmetric,  so the future and past interior actions are the same.

Eqs. \eqref{top and bottom joints at t=0}  cannot be solved exactly.
Both at $ r =\Lambda\rightarrow \infty$ and at  $ r \rightarrow r_-$ the function $r^*$ diverges to $+\infty$,
so we study the behaviour around these points:
\begin{itemize}
\item
Nearby $r \approx r_-$,  we find
\beq
r^* (r_m)  = - \frac{\sqrt{3(\nu^2 -1)}}{\nu^2 +3} \tilde{A} \log \left| r_m - r_- \right| + \tilde{B} + \mathcal{O}(r_m-r_-) \, ,
\label{series expansion rm arond r-}
\eeq
where $\tilde{B} $ is a constant and 
\beq
\tilde{A}=\frac{\sqrt{r_- \Psi(r_-)}}{(r_+-r_-) \sqrt{3 (\nu^2-1)}} > 0 \, .
\eeq
\item Around $ r= \Lambda $,
\beq
r^*_{\Lambda} = \frac{\sqrt{3(\nu^2 -1)}}{\nu^2 +3} \log \Lambda + \tilde{C} + \mathcal{O}(\Lambda^{-1}) \, ,
\label{espa-lambda}
\eeq
where $ \tilde{C} $ is the finite piece of order $ \Lambda^0$ . 
\end{itemize}
Consequently, eq. \eqref{top and bottom joints at t=0} gives: 
\beq
r_m - r_-  \approx \, \Lambda^{- 1/\tilde{A}} \, 
  \exp\left[ \frac{(\tilde{B}-\tilde{C}) (\nu^2 +3)}{\tilde{A} \sqrt{3(\nu^2 -1)}} \right]  \, .
\label{rmin approssimato}
\eeq

{\bf Interior bulk term:}
The interior bulk term can be obtained from eqs. (\ref{bulk12},\ref{bulk3}) and
 has the following logarithmically divergent piece:
 \beq
I^{\rm int}_{\mathcal{V}}  = 2(I_{\mathcal{V}}^1 + I_{\mathcal{V}}^3) 
 = - \frac{l}{4G} (\nu^2 +3) \left[ (r_+ - r_m) r^*_{\Lambda} - \int_{r_m}^{r_+} dr \, r^* (r) \right] \, .
 \label{inte-aaa}
\eeq
The last integral in eq. (\ref{inte-aaa}) is finite, because the function $ r^* (r) $
has integrable singularities in $r \approx r_-, r_+$.
The divergent part of the internal bulk action is:
\beq
I^{\rm int}_{\mathcal{V}}  = - \frac{l}{4G} \sqrt{3(\nu^2 -1)} (r_+ - r_-)  \log \Lambda +\mathcal{O}(\Lambda^0) \, .
\label{internal-bulk-action}
\eeq

{\bf External bulk term:}
We then consider the external part
\beq
I^{\rm ext}_{\mathcal{V}} = 2 \, I_{\mathcal{V}}^2 = - \frac{l}{4G} (\nu^2 +3) \int_{r_+}^{\Lambda} dr \, \le r^*_{\Lambda} - r^*(r) \ri \, .
\label{general formula external bulk rotating case}
\eeq
The behaviour of $r^*(r)$ at large $r$ is:
\beq
r^* (r) = \alpha \log (4 r) + \beta + \frac{\gamma}{r} + \mathcal{O} (r^{-2}) \, ,
\eeq
where 
\bea
\beta &=&
-2 \frac{ \sqrt{r_+ \Psi(r_+)} \log \le \sqrt{r_+} + \sqrt{r_+ - \rho_0} \ri - 
 \sqrt{r_- \Psi(r_-)} \log \le \sqrt{r_-} + \sqrt{r_- - \rho_0} \ri }{(\nu^2 +3)(r_+ - r_-)} \, ,
 \nl
\alpha&=&\frac{\sqrt{3(\nu^2-1)}}{\nu^2 +3}  \, , \qquad
\gamma=\frac{\sqrt{3(\nu^2-1)}}{2(\nu^2+3)} \, (\rho_0-2 r_+-2 r_-)  \, .
\eea
The divergences of  (\ref{general formula external bulk rotating case}) then are
\bea
I^{\rm ext}_{\mathcal{V}} &=&  \frac{l}{4G} (\nu^2 +3) \left[- \alpha  \Lambda + \le \alpha r_+ + \gamma \ri \log \Lambda
\right] + \mathcal{O} (\Lambda^{0}) 
\nl
&=&  -  \Lambda \frac{l}{4G}  \sqrt{3(\nu^2 -1)}+
\frac{l}{8 G}  \sqrt{3(\nu^2 -1)} (\rho_0-2 r_-) (\log \Lambda) 
 + \mathcal{O} (\Lambda^0) \, .
\label{external-bulk-action}
\eea

{\bf Joint terms:}
The action evaluated on the WDW patch has four joint contributions: 
two on the cutoff surface $r = \Lambda$ and two in the region inside the black and white hole. 
They can all be directly evaluated from  eq. (\ref{general formula joints}).
The joint inside the black hole, located at $r = r_{m}$, gives the following contribution:
\beq
\begin{aligned}
I^{r_{m}}_{\mathcal{J}} & = - \, \frac{l}{8 G} \sqrt{r_m \Psi (r_m)} \log \left| \frac{l^2}{A^2} \frac{\le \nu^2 +3 \ri \le r_m - r_- \ri \le r_m - r_+ \ri}{r_m \Psi (r_m)} \right| = \\
& = \frac{l}{8 G} \sqrt{3 \le \nu^2 -1 \ri} \le r_+ - r_- \ri \log \Lambda + \mathcal{O} \le \Lambda^0 \ri \, .
\end{aligned}
\eeq
The joint nearby the cutoff surface gives:
\beq
\begin{aligned}
I^{\Lambda}_{\mathcal{J}} & = \frac{l}{8 G} \sqrt{\Lambda \Psi \le \Lambda \ri} \log \left| \frac{l^2}{A^2} \frac{\le \nu^2 +3 \ri \le \Lambda - r_- \ri \le \Lambda - r_+ \ri}{\Lambda \Psi \le \Lambda \ri} \right| = \\
& =\Lambda  \frac{l}{8 G} \sqrt{3 \le \nu^2 -1 \ri} \log \left| \frac{l^2}{A^2} \frac{\nu^2 +3}{3 \le \nu^2 -1 \ri} \right|  + \mathcal{O} \le \Lambda^0 \ri \, .
\end{aligned}
\eeq
Summing  the contributions of the four joints, we find:
\beq
I^{tot}_{\mathcal{J}} =\Lambda \frac{l}{4 G} \sqrt{3 \le \nu^2 -1 \ri} \log \left| \frac{l^2}{A^2} \frac{\nu^2 +3}{3 \le \nu^2 -1 \ri} \right|  + \frac{l}{4 G} \sqrt{3 \le \nu^2 -1 \ri} \le r_+ - r_- \ri \log \Lambda + \mathcal{O} \le \Lambda^0 \ri \, .
\eeq

{\bf Counterterm:}
The WDW patch is bounded by four codimension-$1$ null surfaces;
they are all the same by symmetry, and so from (\ref{controtermineA}) and (\ref{misu}) we find:
\beq
I_{\rm ct} = \frac{l}{4 G} \int_{r_m}^{\Lambda} dr  \, \frac{\p_r (r \Psi (r) )}{\sqrt{r \, \Psi (r)}} \,
 \log \left| \frac{2 A \tilde{L}}{l^2} \frac{\p_r (r \, \Psi (r))}
{4 \sqrt{r \Psi (r)}} \right| \, .
\eeq
Since $\Psi(r)$ is linear in $r$,
the only divergence  comes from $r = \Lambda$:
\beq
I_{\rm ct} = \Lambda  \frac{l}{4 G} \sqrt{3 \le \nu^2 -1 \ri} \log \left| \frac{\tilde{L}^2 A^2}{l^4} \, 3 \le \nu^2 -1 \ri \right| 
+ \mathcal{O} \le \Lambda^0 \ri \, .
\eeq

{\bf Total action:}
Summing all the contributions, the divergences of the total action are:
\beq
I^{\rm tot}= \frac{l}{4G} \sqrt{3 (\nu^2-1)} \Lambda \left( \log \left( \frac{\tilde{L}^2}{l^2} (\nu^2+3)  \right) -1 \right)
+(\log \Lambda) \frac{l}{4 G}  \sqrt{3(\nu^2-1)} \left( \frac{\rho_0}{2}-r_- \right) \, ,
\label{azione-totale}
\eeq
As in the AdS case, the divergent contribution in the counterterm cancels the dependence
 on the ambiguous normalization constant $A$ appearing in the divergent contribution of the joints.


\subsection{Action of internal region and subregion complexity}

In this section we compute the divergences of the action evaluated on the intersection
 between the WDW patch and the interior of the black and white holes.
 The external part, which  is conjectured to be proportional to the subregion complexity
   of the thermofield double state,
  is then found by subtraction from the total WDW action.
   Again we consider the rotating black hole; the non-rotating case
    is studied in Appendix \ref{appe-exte}.
 The bulk part of the internal action was already computed in (\ref{internal-bulk-action}).

{\bf Joint terms:} 
In the interior of the black hole, there are four contributions of the form (\ref{general formula joints}),
which are all in principle divergent because $f(r_+)=f(r_-)=0$.
Symmetrically, there are other four joints inside the white hole.
As in the AdS case \cite{Agon:2018zso},  due to the signs $\eta=\pm 1$ of each joint,
these divergences will partially cancel each other.

It is useful to introduce the Kruskal coordinates $ (U,V) $ defined for $r>r_-$ as in \cite{Jugeau:2010nq}
\bea
U&=& \sgn (r-r_+) \, e^{b_{*} (r^* (r)-t)}=  \sgn (r-r_+) \,  e^{- b_* u} \, ,  \nl
 V&=&e^{b_{*}(r^* (r)+t)}= e^{b_* v} \, ,
\eea
where 
\beq
b_* = \frac{f'(r_+)}{2} =\frac{(\nu^2 +3)(r_+ - r_-)}{2 \sqrt{r_+ \Psi(r_+)}} \, .
\eeq
These coordinates satisfy the relation
\beq
\log |UV|  =   2b_* r^* (r) =f'(r_+) r^*(r)
\label{log UV coordinates}
\eeq 
which is useful to simplify expressions involving the joints.
Note that, since $r_* \rightarrow - \infty$ when $r \rightarrow r_+$,
the external horizon  corresponds to $U=0$ (black hole horizon for the right boundary)
and $V=0$ (white hole horizon for the right boundary).

\begin{figure}[h]
\centering
\includegraphics[scale=0.7]{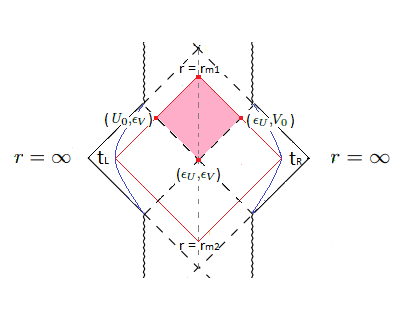}
\caption{Joint terms needed for the action of the black hole interior.}
\label{horizon-joint}
\end{figure}

Let us consider a contribution coming from sums of joints nearby the horizon.
We follow the prescription given in \cite{Agon:2018zso}, introducing the regulators
 $ \varepsilon_U , \varepsilon_V $ to move the joints off the horizon 
 by an infinitesimal quantity.   For instance, if we evaluate the sum of the contributions
  of two terms with the same $V=\varepsilon_V$,
 from eq. (\ref{general formula joints}), we find a term proportional to
\beq
\begin{aligned}
& \log \left| \frac{l^2}{A^2} \frac{f(r_{U_1, \varepsilon_V})}{2R(r_+)} \right| - \log \left| \frac{l^2}{A^2} \frac{f(r_{U_2, \varepsilon_V})}{2R(r_+)} \right|  =  \int_{r_{U_2, \varepsilon_V}}^{r_{U_1, \varepsilon_V}} \frac{dr}{f(r)} f'(r) \approx  f'(r_+) \int_{r_{U_2, \varepsilon_V}}^{r_{U_1, \varepsilon_V}} \frac{dr}{f(r)} = \\
& = f' (r_+) \left[ r^* (r_{U_1, \varepsilon_V}) - r^* (r_{U_2, \varepsilon_V}) \right] =
\log \left| U_1 \varepsilon_V \right| - \log \left|U_2 \varepsilon_V  \right| = \log \left| \frac{U_1}{U_2} \right| \, ,
\end{aligned}
\label{difference of logarithms at horizon}
\eeq
where in the last steps we simplified the result by means of eq. (\ref{log UV coordinates}).

This expression tells us that in the limit $ \varepsilon_V \rightarrow 0, $ the difference of joints at
 the horizon is regular and the divergences coming from each term separately cancel.
We could perform the same trick  exchanging the $ U \leftrightarrow V $ coordinates, 
since the previous manipulations are symmetric under this transformation.
Combining these two results, one can conclude that
\beq
\log \left| \frac{l^2}{A^2} \frac{f(r_{U,V)}}{2 R(r_+)} \right| = \log \left| U V \right| + F(r_+) \, ,
\label{logarithm at the horizon}
\eeq
where the function $ F(r) $ is regular at the horizon and is given by
\beq
F(r) = \log \left| \frac{l^2}{A^2} \frac{f(r)}{2 R(r)} \right| - f'(r_+) r^* (r)   \, .
\label{BIGF}
\eeq

There are four joint contributions inside the black hole and four inside the white hole;
by symmetry they are the same and the total contribution is twice the ones of the black hole:
\beq
\begin{aligned}
I^{\mathrm{int}}_{\mathcal{J}} = & \, - 2 \times  \frac{l}{4 G}  \sqrt{\frac{r_+}{4} \Psi(r_+)} \left[ \log \left|\frac{l^2}{A^2} \frac{f(r_{\epsilon_U,\epsilon_V})}{2 R(r_+)}  \right|-\log \left|\frac{l^2}{A^2} \frac{f(r_{U_0,\epsilon_V})}{2 R(r_+)}  \right|-\log \left|\frac{l^2}{A^2} \frac{f(r_{\epsilon_U,V_0})}{2 R(r_+)}  \right| \right]  
\\ & - 2 \times \frac{l}{4 G} \sqrt{\frac{r_m}{4} \Psi(r_m)} \log \left| \frac{l^2}{A^2} \frac{f(r_m)}{2 R(r_m)} \right|  \, .
\label{joint tot rotating case}
\end{aligned}
\eeq
This expression simplifies to
\bea
I^{\mathrm{int}}_{\mathcal{J}} &=& 
  \frac{l}{4 G} \sqrt{r_+ \Psi(r_+)} \left[ 2 b_* r^*_{\Lambda} 
  + F(r_+) \right]  -  \frac{l}{4 G} \sqrt{r_m \Psi(r_m)} \log \left| \frac{l^2}{A^2} \frac{f(r_{m})}{2 R(r_m)} \right| 
\nl
&=& \frac{l}{2 G} \sqrt{3(\nu^2-1)} (r_+- r_-) \log \Lambda +\mathcal{O}(\Lambda^0) \, .
\eea

{\bf Counterterms:}
The last contribution comes from the counterterm, and possible dependences from the UV cutoff 
can arise only from the $ r=r_m $ endpoint of integration.
However, putting the expansion (\ref{rmin approssimato}) inside the counterterm, 
we find that no  divergent pieces appear.

{\bf Internal and external action:}
Putting together all the terms contributing to the interior action in the rotating case, we find that
the divergent part of the internal action is:
\beq
I^{\rm int} = \frac{l}{4G} \sqrt{3(\nu^2 -1)} (r_+ - r_-) \log \Lambda + \mathcal{O} (\Lambda^0) \, .
\eeq
Subtracting this from eq. (\ref{azione-totale}), we find
the divergences of the external action, which correspond to the subsystem complexity:
\beq
I^{\rm ext}= \frac{l}{4G} \sqrt{3 (\nu^2-1)} \Lambda \left( \log \left( \frac{\tilde{L}^2}{l^2} (\nu^2+3)  \right) -1 \right)
+(\log \Lambda) \frac{l}{4 G}  \sqrt{3(\nu^2-1)} \left( \frac{\rho_0}{2}-r_+ \right) \, .
\label{external-action}
\eeq

\subsection{Comments on regularization}

\begin{figure}[h]
\centering
\includegraphics[scale=0.7]{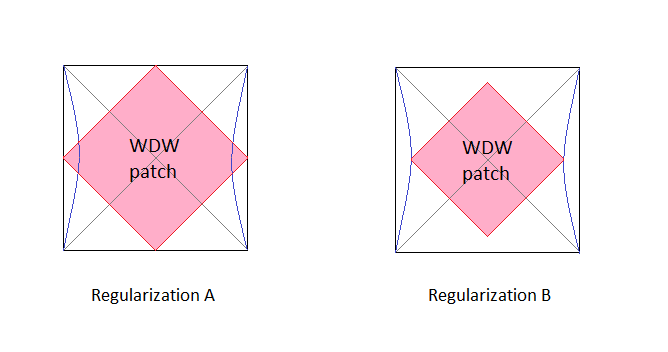}
\caption{Two different regularizations can be
chosen for the action of the WDW patch for black holes in AdS
(here for illustrative purpose we show
the case of non-rotating black hole in asymptotically AdS spacetime).}
\label{2-regs}
\end{figure}

In AdS one can use two different regularizations \cite{Carmi:2016wjl}
for the CA conjecture (see figure \ref{2-regs}): 
\begin{itemize}
\item  one can consider the edge
of the WDW ending on the asymptotic AdS boundary 
(regularization $A$)
\item 
one can consider the edge ending on the regulator surface
(regularization $B$).
\end{itemize}
The two regularizations give the same complexity rate
at large times. In asymptotically AdS spaces,
if one introduces appropriate counterterms
in regularization $A$ one can reproduce the same results
as in regularization $B$ \cite{Kim:2017lrw,Akhavan:2019zax}.

\begin{figure}[h]
\centering
\includegraphics[scale=0.7]{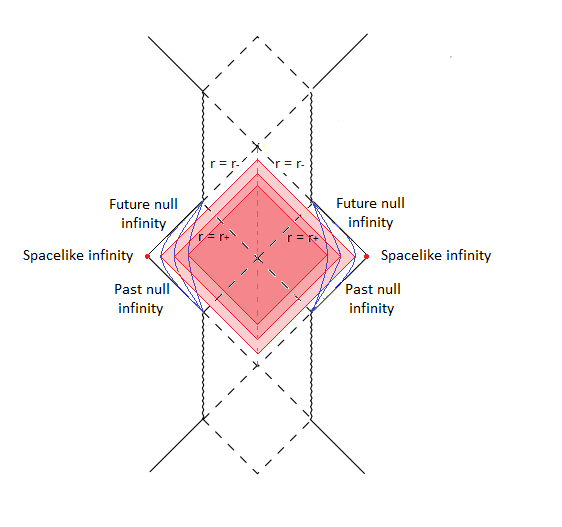}
\caption{In WAdS the causal structure resemble the one
of asymptotically Minkowski spacetime. Regularization $A$
would give a WDW patch with a corner which is located
at the spacelike infinity, and so would give a time-independent complexity.
Moreover, the WDW patch in regularization $B$  covers the entire black hole interior
in the limit of infinite cutoff $\Lambda$.}
\label{cutoff-var}
\end{figure}

In WAdS the structure of the Penrose diagram is radically different from AdS,
and it resembles instead the one of asymptotically Minkowski space:
the right corner of the Penrose diagram corresponds to
$r \rightarrow \infty$ and arbitrary $t$ (spacelike infinity).
The $45$ degrees boundaries correspond to the 
future null infinity and past null infinity (see figure \ref{cutoff-var}).

In all the previous works on the CA conjectures in WAdS 
\cite{Ghodrati:2017roz,Auzzi:2018pbc,Dimov:2019fxp},
 regularization $B$ was implicitly used.  This approach gives 
 the expected result for the complexity rate at late time
 $\dot{\mathcal{C}}_A \propto TS$ in the case of Einstein gravity \cite{Auzzi:2018pbc}.
 We used as well this regularization in the previous section
  to compute subregion action complexity.

It is not straightforward to generalise regularization $A$
to the case of WAdS, because then the corner of the WDW patch
would be located at the spacelike infinity point for all values of the time.
This would give the unphysical result that complexity is time-independent.

If in regularization $B$
we  sent the UV cutoff to infinity,
 we would find that the WDW patch
includes all the interior of the black hole.
This is the same part of the Penrose diagram which gives the 
linear growth of complexity at large time; so sending the cutoff
to infinity is equivalent to sending the time to infinity with finite cutoff, 
and so gives a diverging internal action.
This explains why the action of the internal part of 
the WDW is UV divergent in WAdS, while it is finite in AdS.

\subsection{Comments on counterterms}

Following  \cite{Lehner:2016vdi}, 
we inserted in the gravitational action a counterterm of kind (\ref{counterterm})
which is needed in order to maintain reparameterization invariance
in presence of null boundaries. 
This term is not necessarily unique.

Let us borrow some notation from  \cite{Lehner:2016vdi}.
Let us consider a null hypersurface defined by the function $\Phi(x^\a)=0$.
The hypersurface can be described by parametric equations $x^\mu(\lambda,\theta^A)$,
where $\lambda$ is the affine null parameter and $\theta^A$ is constant on each null 
generator on the surface. 
The vectors:
\beq
k^\mu=\frac{\p x^\mu}{\p \lambda} \, , \qquad
e^\mu_A= \frac{\p x^\mu}{\p \theta^A} \, ,
\eeq
are tangent to the surface; $k^\a$ is the null normal to the surface.
Let us denote by
\beq
\sigma_{AB}= g_{\a \b} e^\a_A e^\b_B
\eeq
the induced metric on the transverse directions $\theta_A$.
Also, one can introduce the following tensor
\beq
B_{AB}= e^\a_A e^\b_B D_\a k_\b \, ,
\eeq
which describes the behaviour of the congruence of null generators.


In principle, as discussed in the Appendix B of \cite{Lehner:2016vdi}, 
in presence of null boundaries
we can also allow for Lagrangians depending on combinations of the Riemann tensor
$\hat{R}_{ABCD}$ computed from the transverse induced metric $ \sigma_{AB} $.
Moreover,  contributions
containing the tensor $ B_{AB}$
are also allowed. A priori  we could have a counterterm of the type
\beq
\mathcal{L}_{\rm ct} (\hat{R}, \hat{R}_{AB}, \hat{R}_{ABCD}, B_{AB}, \Theta) \, ,
\eeq
where we should require that the total action is
 reparametrization-invariant.
Dramatic restrictions arise from the fact that 
we are working in 3 dimensions, which means that the null surfaces are 2-dimensional 
and that the induced metric $ \sigma_{AB} $ is 1-dimensional.
This implies that 
\beq
\hat{R}_{ABCD}=0 \, , \qquad
\hat{R}_{AB} = 0 \, , \qquad
\hat{R} = 0 \, , \qquad
B_{AB} = \frac12 \Theta \, \sigma_{AB} \, ,
\eeq
and then there is no space for curvature terms other than the geodesic expansion parameter
 $ \Theta$, which we already considered for the counterterm (\ref{counterterm}).


\section{Volume}
\label{volume}

\subsection{CV conjecture}

In this section we compute the divergences
 of the volume complexity at $t_b=0$ for the generic rotating black hole
 (the non-rotating case is studied in Appendix \ref{appe-volume},
 where also the finite part is evaluated).
The time dependence of the finite part of the volume was previously studied in  \cite{Auzzi:2018zdu}.

The extremal volume at $t_b=0$ is a constant $t=0$ bulk slice, connecting
the two $t_L=0$ and $t_R=0$ regions on the left and right boundaries.
The RT surface is a line at a constant value of the radial coordinate $ r=r_+$. 
We  denote by $ V(L) $ the volume of the codimension-$1$ extremal surface anchored at the entire left boundary of the spacetime,
 and by $ V(R) $ the corresponding volume for the right boundary.
The symmetry of the problem implies that the subregion complexity on the two boundaries separately is the same, and then
\beq
V^{\rm out} =  V(L) + V(R) = 2 V(L) \, .
\eeq

The volume can be computed directly from the determinant of the induced metric on the $t=0$ slice:
\bea
V(L)  &=& 2 \pi l^2  \int_{r_+}^{\Lambda} dr \, G(r) \, , \nl
G(r) &=&  
 \sqrt{\frac{r \le 3(\nu^2 -1)r +(\nu^2 +3)(r_+ + r_-) -4 \nu \sqrt{r_+ r_- (\nu^2 +3)} \ri}{4 (\nu^2 +3)(r-r_-)(r-r_+)}} \, .
\eea
Let us study the possible divergences of this integral.

Nearby the outer horizon $ r \rightarrow r_+ $,  $ G(r) $ can be approximated by the following expression:
\beq
G(r)  =\frac{g}{\sqrt{r-r_+}} + \mathcal{O} \le  \sqrt{r-r_+} \ri  \, , \quad
g=
 \sqrt{\frac{r_+ \le 4 \nu^2 r_+ + (\nu^2 +3)r_- -4 \nu \sqrt{r_+ r_- (\nu^2 +3)} \ri}{4 (\nu^2 +3)(r_+ - r_-)}} \, .
\eeq
The contribution to the volume coming from the region nearby the outer horizon is:
\beq
2 \pi l^2 \int_{r_+}^{r_+ + \varepsilon} dr \, G(r) \approx
2 \pi l^2 \int_{r_+}^{r_+ + \varepsilon} dr \, \frac{g}{\sqrt{r-r_+}} \approx
4 \pi l^2  g \sqrt{\epsilon}  \, .
\eeq
So there is no divergence nearby the horizon.

At $r \rightarrow \infty$,  $G(r)$ can be expanded as follows:
\beq
G(r) = \sqrt{\frac{3(\nu^2 -1)}{4 (\nu^2 +3)}} + \frac{\nu \le \nu (r_+ + r_-) - \sqrt{r_+ r_- (\nu^2 +3)} \ri}{\sqrt{3(\nu^2 -1)(\nu^2 +3)}} \frac{1}{r} + \mathcal{O} \le \frac{1}{r^2} \ri \, .
\eeq
Upon integration, the first two terms give rise to a linear
 and a logarithmic divergences. The divergence of the volume then is
\beq
V(L) = \pi l^{2} \sqrt{ \frac{3 (\nu^{2} -1)}{\nu^{2}+3} } \Lambda 
+ \frac{32 \pi G l^{2} \nu^{2}}{(\nu^{2}+3)^{3/2} \sqrt{3 (\nu^{2}-1)}} M \log \Lambda 
+ \mathcal{O} \left( \Lambda^{0} \right) \, .
\label{divergenza-volume}
\eeq
The logarithmically divergent term is proportional to the mass $M$.

\subsection{Spacetime volume (CV 2.0)}

It was proposed in  \cite{Couch:2016exn} that complexity is
 dual to the spacetime volume of the WDW patch;
in our case  this is very similar to the action conjecture because 
the bulk term in the action is the integral of a constant.
We can borrow the calculations of the divergences from eqs.
(\ref{internal-bulk-action}) and (\ref{external-bulk-action}):
\bea
V^{\rm int}_{\rm bulk}  &=& 4 \pi l^3  \frac{\sqrt{3(\nu^2 -1)}}{\nu^2+3} (r_+ - r_-)  \log \Lambda +\mathcal{O}(\Lambda^0)\, , \nl
V^{\rm ext}_{\rm bulk}&=& 4  \pi l^3 \frac{\sqrt{3(\nu^2 -1)}}{\nu^2+3} \left(  \Lambda -
 \left(\frac{\rho_0}{2}- r_- \right) \log \Lambda  \right)
 + \mathcal{O} (\Lambda^0) \, .
 \label{CV2}
\eea

\section{Conclusions}
\label{conclu}

{\bf Sub/superadditivity.}
In AdS black holes the internal action  $I^{\rm int}$ at $t_b=0$
 is finite \cite{Alishahiha:2018lfv,Agon:2018zso} and has
a sign which depends on the choice of the counterterm parameter $\tilde{L}$.
In turn, depending on the sign of $I^{\rm int}$, the action subregion complexity
can be sub/superadditive.
Instead, in WAdS$_3$ the interior action $I^{\rm int}$ is always positive 
and independent of the counterterm length scale;
as a consequence, $\mathcal{C}_A$ subregion complexity of the left and right side
of the thermofield double is superadditive.
Moreover, $I^{\rm int}$ is proportional to the product of
 temperature and entropy of the black hole:
\beq
I^{\rm int} = \frac{4 \sqrt{3(\nu^2 -1)}}{\nu^2 +3} \, l \,  TS \, \log \Lambda + \mathcal{O} (\Lambda^0) \, .
\eeq
By construction,  $\mathcal{C}_V$ and $\mathcal{C}_V$ 2.0
are superadditive, with the difference that $\mathcal{C}_V$ saturates superadditivity
(\ref{saturation}) for the left and right side of thermofield double at $t_b=0$, while 
$\mathcal{C}_V$ 2.0 does not, see eq. (\ref{CV2}).

{\bf Structure of divergences.}
For the BTZ black hole, the only divergence in the holographic subregion
complexity is linear in the cutoff $\Lambda$.
In WAdS$_3$, we found that the three versions of holographic subregion complexity
have all a linear and a logarithmic divergence in $\Lambda$.
The coefficient of the linear divergence, as in the BTZ case, can be positive or negative
depending on  the counterterm parameter $\tilde{L}$.
The coefficient of the log divergence is  independent from $\tilde{L}$;
it is instead a function of  the black
holes parameters $r_+,r_-$, or equivalently of $T,J$.
In each of the three versions, the log divergence of the subregion complexity
is proportional to a different quantity:
\begin{itemize}
\item in the CA conjecture, eq. (\ref{external-action}) gives a result proportional to
$K_+=   \frac{\rho_0}{2}-r_+ $, with a positive coefficient;
\item in the CV conjecture, eq. (\ref{divergenza-volume}) gives a term proportional to the mass $M$,
with a positive coefficent;
\item in  CV 2.0, eq. (\ref{CV2}) gives a log divergence proportional to
$K_-=   \frac{\rho_0}{2}-r_- $, with a negative coefficient.
\end{itemize}

{\bf Temperature behaviour.}
For neutral black holes in AdS, subregion $\mathcal{C}_A$ has  different properties 
depending on  the regularization parameter $\tilde{L}$.  For $\tilde{L} \ll l$,
$\mathcal{C}_A$ increases with temperature,
whereas,  for $\tilde{L} \gg l$, $\mathcal{C}_A$ decreases with temperature. 
Instead, for neutral black holes in AdS, subregion $\mathcal{C}_V$ is an increasing powerlike
function of temperature \cite{Chapman:2016hwi} (for AdS$_3$, actually, it is independent of temperature).

\begin{figure}[h]
\centering
\includegraphics[scale=0.7]{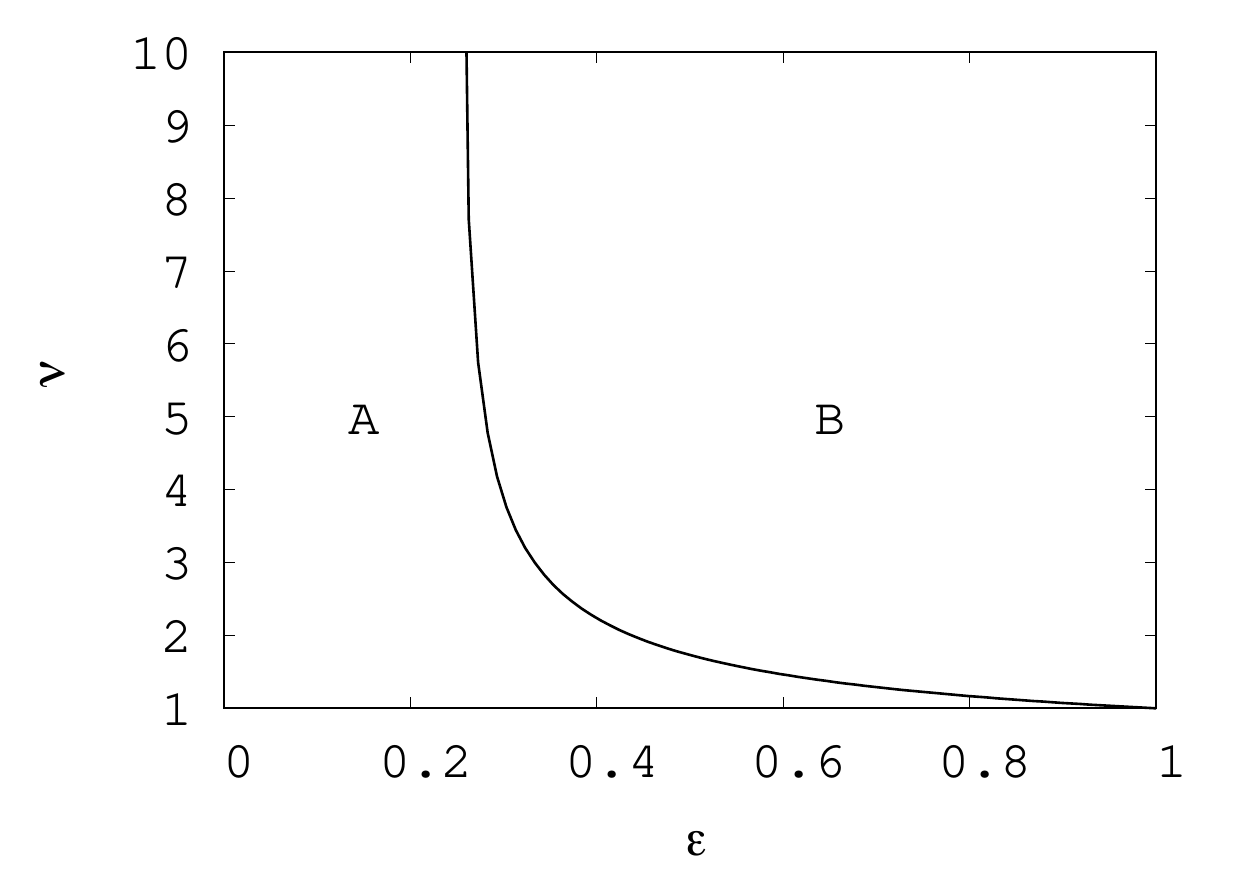}
\caption{Regions A and B in $(\epsilon, \nu)$ plane.
The angular momentum $J$ and the specific heat $C_J$ are negative in region $A$
and positive in region $B$.  $\mathcal{C}_V$ decreases with temperature in 
region $A$ and increases in region $B$.
$\mathcal{C}_A$ increases with temperature in 
region $A$ and decreases in region $B$.
}
\label{regionsAB}
\end{figure}

In WAdS$_3$, the leading dependence on temperature of the subsystem complexity
is in the $\log \Lambda $ terms. To this purpose we introduce 
\beq
C_J=\frac{\p M}{\p T} \Big|_J \, , \qquad C_+=\frac{\p K_+}{\p T} \Big|_J \, , \qquad
C_-=\frac{\p K_-}{\p T} \Big|_J \, ,
\label{tante-C}
\eeq
which are explicitly computed in Appendix \ref{appe-temperature}. $C_J$
is the specific heat at constant $J$.
We note that the scale $r_+$ factorises from the quantities (\ref{tante-C}), so we introduce
\beq
\epsilon=r_-/r_+  , \qquad
 0 \leq \epsilon < 1 \, ,
\eeq
and we study the sign of (\ref{tante-C}) as a function of $(\epsilon,\nu)$.
Let us define (see figure \ref{regionsAB})
\bea
{\rm Region \,\,\, A:} &&  \qquad  0 < \epsilon< \frac{\nu^2+3}{4 \nu^2} \equiv \epsilon_c(\nu) \\
{\rm Region \,\,\, B:} &&  \qquad   \epsilon_c(\nu) < \epsilon <1 \, .
\label{AABB}
\eea

The angular momentum $J$ is negative in region A and positive in region B,
while it vanishes along the two curves  $\epsilon=0$ and $ \epsilon= \epsilon_c(\nu)$.
It is interesting that the quantities $(C_J,C_+)$
change sign in regions $A,B$: 
\begin{itemize}
\item $C_J$ is positive in region $B$ and negative in region $A$;
\item $C_+$ is negative in region $B$ and positive in region $A$.
\end{itemize}
As a consequence, in the region where $C_J>0$,
subregion $\mathcal{C}_V$   increases with temperature (at constant $J$),
while $\mathcal{C}_A$ decreases. In the thermodynamically unstable region
where $C_J<0$,  subregion $\mathcal{C}_V$  decreases with temperature
while $\mathcal{C}_A$ increases.

\begin{figure}[h]
\centering
\includegraphics[scale=0.7]{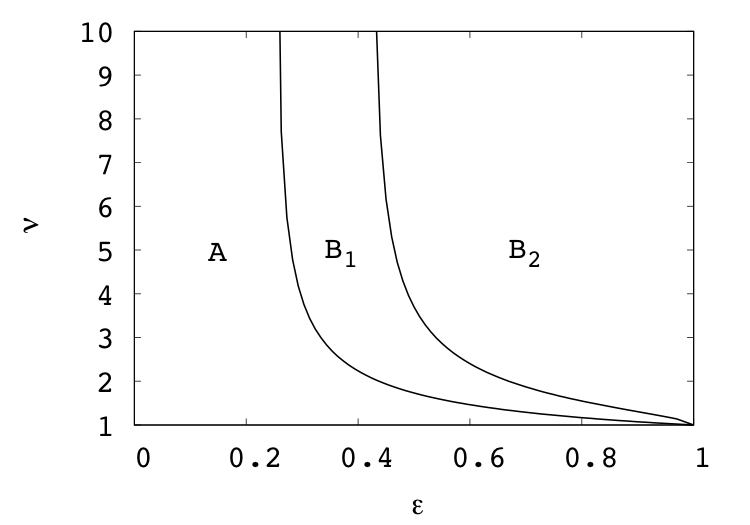}
\caption{Regions of $(\epsilon,\nu)$ with different temperature behaviour of subregion $\mathcal{C}_V^{\rm 2.0}$: it
decreases with temperature at constant $J$
in regions $A$ and $B_2$, while it increases in region $B_1$.}
\label{regionsABC}
\end{figure}

The behaviour of spacetime volume $\mathcal{C}_V^{\rm 2.0}$ with temperature is shown in figure 
 \ref{regionsABC}, where the region $A$ is still the one defined
 in eq. (\ref{AABB}) and region $B$ splits into regions $B_1$ and $B_2$.
Subregion spacetime volume decreases with temperature at constant $J$
in regions $A$ and $B_2$, while it increases in region $B_1$.
Note that the three quantities $(C_J,C_+,C_-)$ all diverge
along the curves $\epsilon=0$ and $ \epsilon= \epsilon_c(\nu)$.

\bigskip

In this work we find that  in WAdS$_3$ the properties of $\mathcal{C}_A$ are somehow 
more robust compared to the AdS case, because they do not depend  on the value of $\tilde{L}$:
$\mathcal{C}_A$ is universally superadditive for the $L,R$ states of the TD state and
the behaviour of subregion complexity with temperature is correlated with the specific heat.

On general grounds,   $\mathcal{C}_B$ is expected to decrease with $T$ \cite{Agon:2018zso}.
The sub/superadditivity properties of subregion  $\mathcal{C}_P$ and $\mathcal{C}_B$
are not firmly established and further studies are desirable.
According to  \cite{Agon:2018zso},  $\mathcal{C}_P$ should be subadditive
for the $L$ and  $R$ factors of $TD$, while $\mathcal{C}_B$ should be
superadditive (see however \cite{Caceres:2018blh} 
for an independent discussion which suggests that purification complexity
might be neither sub nor superadditive).  
Our findings for $\mathcal{C}_A$ seem to match these expectations for $\mathcal{C}_B$
in the thermodynamically stable region with $C_J>0$.
On the contrary, $\mathcal{C}_V$ seems to match with  $\mathcal{C}_B$
in the unstable region with $C_J<0$.

The physical meaning of the scale $\tilde{L}$ in the CA conjecture
is still rather obscure. In AdS many properties of 
subregion complexity are sensitive to this parameter.
It is possible that $\tilde{L}$  is somehow related
to the regularization procedure needed to define
complexity in quantum field theory; it would be interesting
to study this issue. In WAdS$_3$ instead the scale $\tilde{L}$
has less influence on the physical properties of holographic complexity.

 The properties of purification and basis complexity deserve further investigation,
in order to understand if some universal features can emerge which
can be matched with general features of holographic complexity.
It would be also interesting to search for other measures of mixed state quantum 
complexity that may have different properties.


 \section*{Acknowledgments}

AM acknowledges the financial support from a Fellowship from the ICTP Programme for TRIL, Trieste, 
Italy and  the hospitality of UCSC, Brescia, Italy, while this work was in progress.


\section*{Appendix}
\addtocontents{toc}{\protect\setcounter{tocdepth}{1}}
\appendix

\section{Non-rotating case}
\label{appe-non-rot}

In this appendix we consider in detail the non-rotating case
with $r_+=r_h$ and $r_-=0$ and we check that the divergences
of complexity reproduce the appropriate limit of the rotating case.

\subsection{Total action}
\label{appe-total}

Summing eqs (\ref{bulk12},\ref{bulk3}) the bulk action is
\beq
\begin{aligned}
I_{\mathcal{V}}^{\mathrm{tot}} = \frac{\mathcal{I}}{2G} \int_{0}^{\Lambda} dr \, (r^*_{\Lambda}- r^* (r)) = - \frac{l}{4G} (\nu^2 +3) \Lambda  r^*_{\Lambda} + \frac{l}{4G} (\nu^2 +3) \int_{_0}^{\Lambda} dr \, r^* (r)  \, .
\label{total bulk action before critical time}
\end{aligned}
\eeq
The GHY term  (\ref{gibb-hawk})  gives \cite{Auzzi:2018pbc}:
\beq
I_{\mathcal{B}}= - \frac{(\nu^2 +3)l}{4G} (2 \varepsilon_0 - r_h) (r^*_{\Lambda} - r^*(\varepsilon_0)) = \frac{(\nu^2 +3)l}{4G} r_h \, 
\le r^*_{\Lambda} - r^* (0) \ri  \, , 
\eeq
where in the last step we performed the limit $ \varepsilon_0 \rightarrow 0 $.
The expression is divergent after sending $ \Lambda \rightarrow \infty $ due to the behaviour at infinity of the tortoise coordinate.

At $t_b=0$, the joints of the WDW patch are located at both the IR and UV cutoffs.
The former vanish as already observed, while the latter give the only non-vanishing expression.
If we conventionally decide to take the flow of time in the bulk as increasing when going upwards, these joints take a negative sign $ \eta=-1 $ in eq. (\ref{general formula joints}) and we obtain
\beq
I_{\mathcal{J}} = 2 \times \frac{l}{4 G} \sqrt{\frac{\Lambda}{4} \Psi (\Lambda)} \log \left|\frac{l^2}{A^2} \frac{f(\Lambda)}{2 R (\Lambda)}  \right| =
\frac{l}{4 G} \sqrt{\Lambda \Psi (\Lambda)} \log \left|\frac{l^2}{A^2} \frac{(\nu^2 +3)(\Lambda - r_h)}{\Psi (\Lambda)}  \right|
 \, .
\eeq 
Finally, we have to add the counterterm which renders the action diffeomorphism-invariant:
\beq
I_{\rm ct}  = 4 \times
\frac{l}{4 G} \int_{\varepsilon_0}^{\Lambda} dr \, \frac{6(\nu^2 -1)r + (\nu^2 +3)r_h}{4 \sqrt{ r \Psi (r)}} \, \log \left| \frac{A \tilde{L}}{2 l^2} \frac{6(\nu^2 -1)r + (\nu^2 +3) r_h}{\sqrt{r \Psi (r)}}  \right| \, .
\eeq
The integration can be done analytically and we can also perform  the usual limit $ \varepsilon_0 \rightarrow 0 , $ finding
\begin{align}
I_{\mathrm{ct}}=\frac{l}{4G}&\left[\frac{2(\nu^2+3)r_h}{\sqrt{3(\nu^2-1)}} \arctan \left(\frac{\sqrt{3(\nu^2-1)\Lambda}}{\sqrt{(\nu^2+3)r_h+3(\nu^2-1)\Lambda}}\right) \right. \notag\\&\left.-\sqrt{\Lambda \Psi (\Lambda)}\log\left|\frac{4l^4}{A^2\tilde{L}^2}\frac{\Lambda \Psi (\Lambda)}{((\nu^2+3)r_h+6(\nu^2-1)\Lambda)^2}\right| \right] \, .
\end{align}
Putting all these results together we obtain the expression for the total action in the WDW patch
\beq
\begin{aligned}
I^{\rm tot}  &=   \frac{l}{4G} (\nu^2 +3) \int_{0}^{\Lambda} dr \, r^* (r) 
- \frac{l}{4G} (\nu^2 +3) \Lambda  r^*_{\Lambda} + \frac{(\nu^2 +3)l}{4G} r_h \, \le r^*_{\Lambda} - r^* (0) \ri \\
& + \frac{l}{2G} \frac{(\nu^2 +3)r_h}{\sqrt{3(\nu^2 -1)}} \arctan \le \frac{\sqrt{3(\nu^2 -1)\Lambda}}{\sqrt{(\nu^2 +3)r_h + 3 (\nu^2 -1)\Lambda}} \ri \\
& + \frac{l}{4G} \sqrt{\Lambda \Psi (\Lambda)} \log \left| \frac{\tilde{L}^2}{4 l^2} \frac{(\nu^2 +3)(\Lambda-r_h)\left[(\nu^2 +3)r_h +6(\nu^2 -1)\Lambda \right]^2}{\Lambda \Psi^2(\Lambda)} \right| \, .
\label{total action before t_C}
\end{aligned}
\eeq

The divergent parts of the total complexity are:
\beq
I^{\rm tot} = 
\frac{l}{4G} 
\sqrt{3(\nu^2 -1)} \Lambda \left( \log \left| \frac{\tilde{L}^2}{l^2} (\nu^2 +3) \right| -1  \right) 
- \frac{l}{8G} \frac{\nu^2 +3}{\sqrt{3(\nu^2 -1)}} \, r_h \log \Lambda
 + \mathcal{O} (\Lambda^0) \, . 
 \label{UV behaviour total action}
\eeq
This reproduces eq. (\ref{azione-totale}) in the $r_- \rightarrow 0$ limit.

\subsection{External action}
\label{appe-exte}

The bulk and the counterterm action can be obtained in the same way as in Appendix \ref{appe-total}:
\beq
I^{\rm out}_{\mathcal{V}} = - \frac{l}{4G} \sqrt{3(\nu^2 -1)} \Lambda - \frac{l}{8G} \frac{7 \nu^2 -3}{\sqrt{3(\nu^2 -1)}} \, r_h \log \Lambda + \frac{l}{4G} (\nu^2 +3) r_h \, r^*_{\Lambda}
+ \mathcal{O} (\Lambda^0) \, , 
\eeq
\beq
I^{\rm out}_{\rm ct} = - \frac{l}{4G} \sqrt{\Lambda \psi (\Lambda)} \log \left| \frac{4l^4}{A^2 \tilde{L}^2}  \frac{\Lambda \Psi(\Lambda)}{\left[ 6(\nu^2 -1)\Lambda + (\nu^2 +3) r_h \right]^2} \right| + \mathcal{O} (\Lambda^0)
 \, .
\eeq
There is no spacelike or timelike boundary, so there is no contribution from the GHY term.

As in the rotating case, we need to be careful
 with the regularization of the joints at the horizon; we use again 
 the same method as in \cite{Agon:2018zso}.
From (\ref{general formula joints}) in this situation, we find
\begin{align}
I^{\mathrm{out}}_{\mathcal{J}} =  & - \frac{ l}{4 G} \sqrt{r_h \Psi(r_h)}  \left[-\log \left|\frac{l^2}{A^2} \frac{f(r_{\epsilon_U,\epsilon_V})}{2R(r_h)}  \right|+\log \left|\frac{l^2}{A^2} \frac{f(r_{U_0, \epsilon_V})}{2R(r_h)}  \right|
+\log \left|\frac{l^2}{A^2} \frac{f(r_{\epsilon_U, V_0})}{2R(r_h)}  \right| \right] \notag\\& 
+ \frac{l}{4 G} \sqrt{\Lambda
\Psi(\Lambda)}\log \left| \frac{l^2}{A^2} \frac{f(\Lambda))}{2R(\Lambda)} \right| \, .
\end{align}
In this case it is convenient to add and subtract the joint term $\frac{ l}{2 G} \nu r_h \log \left|\frac{l^2}{A^2} \frac{f(r_{\epsilon_U,\epsilon_V})}{2\nu r_h}  \right|$ and to use the relation (\ref{difference of logarithms at horizon}) to get
\beq
\begin{aligned}
I^{\mathrm{out}}_{\mathcal{J}} = &  -\frac{l}{2 G} \nu r_h \left[ \log(U_0 V_0)+\log \left|\frac{l^2}{A^2} \frac{f(r_{\epsilon_U,\epsilon_V})}{2\nu r_h}  \right|-\log(\epsilon_U\epsilon_V)  \right]  \\ & + \frac{l}{4 G} \sqrt{\Lambda
\Psi(\Lambda)}\log \left| \frac{l^2}{A^2} \frac{(\nu^2 +3)(\Lambda - r_h)}{\Psi(\Lambda)} \right|  \, .
\end{aligned}
\eeq
Finally, the expression simplifies by means of eqs. (\ref{log UV coordinates}) and (\ref{logarithm at the horizon}):
\begin{align}
I^{\mathrm{out}}_{\mathcal{J}} = &-\frac{ l}{2 G}\left[ \nu r_h \left(\frac{\nu^2 +3}{2 \nu} r^{*}_{\Lambda} +F(r_h)  \right) - \frac12 \sqrt{\Lambda
\Psi(\Lambda)}\log \left| \frac{l^2}{A^2} \frac{(\nu^2 +3)(\Lambda - r_h)}{\Psi(\Lambda)} \right| \right] \, .
\end{align}
The function $ F(r) $, which can be obtained from eq. (\ref{BIGF}), is finite
 and it is not needed to find the divergence structure.
Adding all the terms outside the horizon, we finally obtain
\beq
I^{\rm out}  =  \frac{l}{4G} \sqrt{3(\nu^2 -1)} \Lambda  \le \log \left| \frac{\tilde{L}^2}{ l^2} (\nu^2 +3)  \right| -1 \ri 
 - \frac{l}{8G} \frac{7 \nu^2 -3}{\sqrt{3(\nu^2 -1)}} \, r_h \log \Lambda + \mathcal{O} (\Lambda^0) \, .
 \label{final outside action}
\eeq
This results reproduces eq. (\ref{external-action})  in the $r_- \rightarrow 0$ limit.

\subsection{Volume} 
\label{appe-volume}

The volume is given by the induced metric computed from the non-rotating metric:
\beq
V (L)= \int_0^{2 \pi} d \theta \int_{r_h}^{\Lambda} dr \, \sqrt{h} =
2 \pi l^2 \int_{r_h}^{\Lambda} dr \,
\sqrt{\frac{3(\nu^2 -1)r + (\nu^2 +3)r_h}{4 (\nu^2 +3)(r-r_h)}} \, .
\eeq
We introduce the coordinate $ R= r/r_h $ and we obtain
\beq
V (L)= 2 \pi l^2 r_h \int_{1}^{\Lambda/r_h} dR \, \sqrt{\frac{3(\nu^2 -1)R + (\nu^2 +3)}{4 (\nu^2 +3)(R-1)}} \, .
\eeq
This expression can be analytically solved, giving
\beq
\begin{aligned}
V (L) & = 2 \pi l^2 r_h \left[ \sqrt{\frac{(\nu^2 +3)+ 3R(\nu^2 -1)}{\nu^2 +3}} \frac{\sqrt{R-1}}{2}  + \right. \\
& \left.  + \frac{2 \nu^2 \log\le \frac{\sqrt{3 (\nu^2 -1)(R-1)} + \sqrt{3+ \nu^2 + 3R(\nu^2 -1)}}{2 \nu}  \ri }{\sqrt{3 (\nu^2 -1)(\nu^2 +3)}}  \right]_{R=1}^{R=\Lambda/r_h} \, .
\end{aligned}
\eeq
This gives the following result:
\beq
\begin{aligned}
V (L) & = \pi l^2 \sqrt{\frac{3(\nu^2 -1)}{\nu^2 +3}} \Lambda +
\frac{2 \pi l^2 \nu^2 r_h}{\sqrt{3(\nu^2 -1)(\nu^2 +3)}} \, \log \le \frac{\Lambda}{r_h} \ri + \\
& + \pi l^2 r_h \frac{(3- \nu^2) + 2 \nu^2 \log \left[ \frac{3(\nu^2 -1)}{\nu^2} \right]}{2 \sqrt{3(\nu^2 -1)(\nu^2 +3)}} + \mathcal{O} (\Lambda^{-1}) \, . 
\end{aligned}
\eeq
The divergent parts of this expression reproduce eq. (\ref{divergenza-volume}) in the $r_- \rightarrow 0$ limit.

\section{Subsystem complexity and temperature}
\label{appe-temperature}

Let us compute the temperature dependence of $M$ at constant $J$, which is the specific heat at constant $J$:
\beq
C_J=\frac{\p M}{\p T} \Big|_J=\frac{\p M}{\p r_+} \frac{\p r_+}{\p T} + 
\frac{\p M}{\p r_-} \frac{\p r_-}{\p T} \, .
\eeq
The quantities $\frac{\p r_+}{\p T}$ and  $\frac{\p r_-}{\p T}$
can be computed from the inverse of the matrix 
\beq
\begin{pmatrix}
 \frac{\p T}{\p r_+}&  \frac{\p T}{\p r_-} \\
 \frac{\p J}{\p r_+}&  \frac{\p J}{\p r_-}
\end{pmatrix} \, ,
\eeq
 which can be directly calculated from eqs. (\ref{TT}) and (\ref{JJ}).
 This gives (here $\epsilon=r_-/r_+ $ ):
\beq
 C_J=\frac{\pi l r_+}{4 G} \, 
 \frac{ \nu  (\epsilon -1) \left(\epsilon  \left(-3 \nu ^2+2 \nu 
   \sqrt{\left(\nu ^2+3\right) \epsilon }+3\right)-2 \nu  \sqrt{\left(\nu ^2+3\right)
   \epsilon }\right)}{ \epsilon  \left(\nu ^2 (4 \epsilon -1)-3\right)} \, .
\eeq
 $C_J$   is negative for
 $ 0< \epsilon< \frac{\nu^2+3}{4 \nu^2}$ and positive for $   \frac{\nu^2+3}{4 \nu^2} < \epsilon <1$.
For $\epsilon=0$  and $\epsilon=\frac{\nu^2+3}{4 \nu^2}$, $C_J$ is diverging and there is a second order phase transition,
similar to the one which occurs for Kerr and Reissner-Nordstr\"om black holes in flat spacetime
\cite{Davies:1978mf}.

With a similar method, one can compute the temperature dependence of
 $K_+$ and $K_-$. The result is:
\beq
\frac{\p K_+}{\p T} \Big|_J=\frac{\hat{a}}{\hat{b}} \, , \qquad \frac{\p K_-}{\p T} \Big|_J=\frac{\hat{c}}{\hat{b}} \, ,
\eeq
where
\bea
\hat{a}&=&2 \pi  l r_+ \left(\sqrt{\left(\nu ^2+3\right) r_+^2 \epsilon }-2 \nu  r_+\right){}^2
   \left(\nu  \left(\nu ^2 ((\epsilon -18) \epsilon -7) 
      +3 \epsilon  (\epsilon
   +6)+3\right) 
   \right.
  \nl && \left. 
\sqrt{\left(\nu ^2+3\right) r_+^2 \epsilon }-r_+ \epsilon  \left(-31 \nu
   ^4+6 \nu ^2+\left(\nu ^2+3\right)^2 \epsilon +9\right)\right) \, ,
\eea
\bea
\hat{b}&=&3 \left(\nu ^2-1\right) \sqrt{\left(\nu ^2+3\right) r_+^2 \epsilon } \left(4 \nu 
   \sqrt{\left(\nu ^2+3\right) r_+^2 \epsilon }+\left(\nu ^2+3\right) r_+ (-\epsilon
   -1)\right) \nl
  &&  \left(2 \nu  (\epsilon +1) \sqrt{\left(\nu ^2+3\right) r_+^2 \epsilon
   }-\left(5 \nu ^2+3\right) r_+ \epsilon \right) \, .
\eea
\bea
\hat{c}&=&2 \pi  l r_+ \left(\sqrt{\left(\nu ^2+3\right) r_+^2 \epsilon }-2 \nu  r_+\right){}^2
   \left(\nu  \left(\nu ^2 (\epsilon  (7 \epsilon +18)-1) 
     -3 (\epsilon  (\epsilon
   +6)+1)\right)
   \right. 
\nl
&&   \left. 
 \sqrt{\left(\nu ^2+3\right) r_+^2 \epsilon }
+r_+ \epsilon 
   \left(\left(\nu ^2+3\right)^2+\left(-31 \nu ^4+6 \nu ^2+9\right) \epsilon
   \right)\right) \, .
\eea

\end{document}